\documentstyle[12pt]{article}
\newcommand{\eqn}[2]{\begin{equation} #2 \label{#1} \end{equation}}
\newcommand{\eqna}[2]{\begin{eqnarray} #2 \label{#1} \end{eqnarray}}
\renewcommand{\theequation}{\thesection.\arabic{equation}}
\newcommand{\newsec}[1]{\setcounter{equation}{0} \section{#1} }
\newcommand{\newapp}[1]{\setcounter{equation}{0}
  \renewcommand{\theequation}{\Alph{section}.\arabic{equation}}
  \renewcommand{\thesection}{\Alph{section}} \section{\bf #1} }

\def\al{\alpha} \def\be{\beta} \def\ga{\gamma}  
\def\de{\delta}  \def\De{\Delta}

\def\la{\lambda} \def\La{\Lambda} \def\rh{\rho} \def\si{\sigma}
    
\def\ph{\phi}     
     
   \def\cO{{\cal O}}
\def\cU{{\cal U}}  \def\cW{{\cal W}} 
\def\cP{{\cal P}}
\def\ie{{\it i.e.\ }}
\def\eg{{\it e.g.\ }}
\def\amsyes{y }
\ifx\answ\amsyes
\expandafter\chardef\csname pre amssym.def at\endcsname=\the\catcode`\@
\catcode`\@=11

\def\undefine#1{\let#1\undefined}
\def\newsymbol#1#2#3#4#5{\let\next@\relax
 \ifnum#2=\@ne\let\next@\msafam@\else
 \ifnum#2=\tw@\let\next@\msbfam@\fi\fi
 \mathchardef#1="#3\next@#4#5}
\def\mathhexbox@#1#2#3{\relax
 \ifmmode\mathpalette{}{\m@th\mathchar"#1#2#3}%
 \else\leavevmode\hbox{$\m@th\mathchar"#1#2#3$}\fi}
\def\hexnumber@#1{\ifcase#1 0\or 1\or 2\or 3\or 4\or 5\or 6\or 7\or 8\or
 9\or A\or B\or C\or D\or E\or F\fi}

\font\tenmsa=msam10
\font\sevenmsa=msam7
\font\fivemsa=msam5
\newfam\msafam
\textfont\msafam=\tenmsa
\scriptfont\msafam=\sevenmsa
\scriptscriptfont\msafam=\fivemsa
\edef\msafam@{\hexnumber@\msafam}
\mathchardef\dabar@"0\msafam@39
\def\dashrightarrow{\mathrel{\dabar@\dabar@\mathchar"0\msafam@4B}}
\def\dashleftarrow{\mathrel{\mathchar"0\msafam@4C\dabar@\dabar@}}

\def\ulcorner{\delimiter"4\msafam@70\msafam@70 }
\def\urcorner{\delimiter"5\msafam@71\msafam@71 }
\def\llcorner{\delimiter"4\msafam@78\msafam@78 }
\def\lrcorner{\delimiter"5\msafam@79\msafam@79 }
\def\yen{{\mathhexbox@\msafam@55 }}
\def\checkmark{{\mathhexbox@\msafam@58 }}
\def\circledR{{\mathhexbox@\msafam@72 }}
\def\maltese{{\mathhexbox@\msafam@7A }}

\font\tenmsb=msbm10 scaled\magstep1   
\font\sevenmsb=msbm7 scaled\magstep1
\font\fivemsb=msbm5 scaled\magstep1
\newfam\msbfam
\textfont\msbfam=\tenmsb
\scriptfont\msbfam=\sevenmsb
\scriptscriptfont\msbfam=\fivemsb
\edef\msbfam@{\hexnumber@\msbfam}

\font\teneufm=eufm10 scaled\magstep1   
\font\seveneufm=eufm7 scaled\magstep1
\font\fiveeufm=eufm5 scaled\magstep1
\newfam\eufmfam
\textfont\eufmfam=\teneufm
\scriptfont\eufmfam=\seveneufm
\scriptscriptfont\eufmfam=\fiveeufm

\catcode`\@=\csname pre amssym.def at\endcsname

\def\CC{{\Bbb C}}

\def\NN{{\Bbb N}}
\def\QQ{{\Bbb Q}}
\def\bfg{{\frak g}}
\def\bfh{{\frak h}}
\def\bft{{\frak h}}

\def\bfnp{{\frak n}_+}
\def\bfnm{{\frak n}_-}
\def\hg{{\widehat{\frak g}}}
\else

\def\CC{{I\!\!\!\!C}}
\def\NN{{I\!\!N}}
\def\QQ{{I\!\!\!\!Q}}
\def\bfg{{\bf g}}
\def\bfh{{\bf h}}
\def\bfnm{{\bf n}_-}
\def\bfnp{{\bf n}_+}
\def\hg{\hat{\bf g}}
\def\bft{{\bf t}}

\fi
\def\wW{{\widehat{W}}}
\def\half{{\textstyle{1\over2}}}
\def\Lap{\La^{(+)}} \def\Lam{\La^{(-)}}
\def\rmH{{\rm H}}
\def\rmHtw{\rmH_{\rm tw}}
\def\gh{{\rm gh}}
\def\AnM#1{Ann.\ Math.\ {\bf #1} }
\def\CMP#1{Comm.\ Math.\ Phys.\ {\bf #1} }
\def\JGP#1{J.\ Geom.\ Phys.\ {\bf #1} }
\def\LMP#1{Lett.\ Math.\ Phys.\ {\bf #1} }
\def\NPB#1{Nucl.\ Phys.\ {\bf B#1} }
\def\PLB#1{Phys.\ Lett.\ {\bf B#1} }
\def\PRep#1{Phys.\ Rep.\ {\bf #1} }
\def\PTP#1{Prog.\ Theor.\ Phys.\ Suppl.\ {\bf #1} }
\def\TAMS#1{Trans.\ Amer.\ Math.\ Soc.\ {\bf #1} }
\def\UMN#1{Usp.\ Mat.\ Nauk\ {\bf #1} }
\def\hepth#1{{\tt hep-th/{#1}}}
\begin{document}
\hsize37truepc\vsize61truepc
\hoffset=-.5truein
\setlength{\baselineskip}{17pt plus 1pt minus 1pt}
\textheight=23cm
\voffset-2.5truecm
\begin{titlepage}

\vphantom{0}\vskip1.1truein
\begin{center}
{\bf ON THE $\cW$-GRAVITY SPECTRUM AND ITS $G$-STRUCTURE
\footnote[5]{Published in the proceedings of the workshop ``Strings, Conformal
Models and Topological Field Theory,'' eds.~L.~Baulieu et al.,
Carg\`ese, May~'93, (Plenum Press, New York, 1995), pp. 59-70.}}
\vskip.65truein

Peter Bouwknegt$^1$, Jim McCarthy$^2$ and Krzysztof Pilch$^1$\\
\vskip.1truein
$^1$ Department of Physics and Astronomy\\
University of Southern California\\
Los Angeles, CA~90089-0484, USA
\vskip.1truein
$^2$ Department of Physics and Mathematical Physics\\
University of Adelaide\\
Adelaide, SA~5005, Australia
\end{center}
\vskip .65truein
\noindent
Abstract: We present results for the BRST cohomology of $\cW[\bfg]$ minimal
models coupled to $\cW[\bfg]$ gravity, as well as scalar fields coupled
to $\cW[\bfg]$ gravity. In the latter case we explore an intricate
relation to the (twisted) $\bfg$ cohomology of a product of two twisted
Fock modules.

\vspace{5truecm}

\begin{flushleft}
{USC-93/27}\\
{ADP-93-226/M21}\\
{\tt hep-th/9311137}
\end{flushleft}

\end{titlepage}
\newsec{Introduction}

The BRST quantization \index{BRST quantization}
of two dimensional $\cW[\bfg]$ gravity \index{W-gravity} coupled to
$\cW[\bfg]$ matter poses the interesting mathematical problem of computing
the semi-infinite cohomology of a $\cW$-algebra \index{W-algebra}
with values in a tensor product
of two (positive energy) $\cW$-modules. In this note we study this
cohomology both for free scalar fields as well as for $\cW$ minimal models
\index{minimal model}
coupled to $\cW$-gravity, \ie we study the cohomology of the tensor
products of
two Fock spaces \index{Fock space}
at irrational $\al_+^2$, and of an irreducible $\cW$-module
with a Fock space. In these cases we give the complete results for the
cohomologies. The work described in this paper is an extension of
\cite{BMPc,BMPd}, where we presented results for the case in which the
`Liouville' momentum takes values in one specific Weyl chamber.
We refer to \cite{BMPc} for further references on the subject.

Strictly speaking, the relevant BRST operator has only been shown
to exist, by explicit construction, for $\cW_3 \equiv
\cW[sl(3)]$ \cite{TM,BLNWa}.
However, since our analysis is insensitive to the specific form of the
BRST operator,
pending the existence proof we have formulated
our results for arbitrary simple, simply-laced Lie algebras $\bfg$.

For irrational $\al_+^2$,
it turns out that there is an intimate connection between the
$\cW[\bfg]$ cohomology \index{W-cohomology}
of a tensor product of two $\cW$ Fock spaces
and the (twisted) $\bfg$ cohomology \index{G-cohomology}
of the product of two twisted
$\bfg$ Fock spaces, which is the finite-dimensional analogue of a
$G/G$ coset model. \index{G/G-model}
Closely related observations have been made in \cite{Sad,Son}.

This note is organized as follows. In section 2 we discuss
the (twisted) $\bfg$ cohomology of the product of two twisted
Fock spaces. In appendix A we give the complete result for $\bfg\cong
sl(2)$ and $sl(3)$.
Our results in this section mainly serve the purpose to formulate the
results for the $\cW$-cohomology through the correspondence alluded to
above, but we believe they are also interesting in their own right.
In section 3 we consider the $\cW[\bfg]$ cohomology
of a tensor product of two Fock spaces and explain the correspondence
with the (twisted) $\bfg$ cohomology. Finally, in section 4, we present
a complete result for the $\cW[\bfg]$ minimal models coupled to
$\cW[\bfg]$ gravity. At the end we included a table of some of the states
for explicitness. We compare
our results to previously obtained results, in particular
those of \cite{BLNWb,Pope}, and find complete agreement.


\newsec{$G$-cohomology of a product of two twisted Fock spaces}

Let $\bfg$ be a finite-dimensional simple Lie algebra. \index{Lie algebra}
Fix a triangular
decomposition $\bfg \cong \bfnm \oplus \bfh \oplus \bfnp$, and a
corresponding Chevalley basis $\{e_{-\al}, h_i, e_{\al}\},\,
\al\in\De_+,\, i=1,\ldots, {\rm rank\,}\bfg$. For any
$\bfg$-module $V$ in the BGG-category
$\cO$ \cite{BGG} (loosely speaking, the category of modules with weights
bounded from above),
we can consider its `twisted' cohomology \index{twisted cohomology}
${\rmH}^i_{\rm tw}(\bfg,V)$, which is the finite-dimensional
analogue of the so-called `semi-infinite' cohomology introduced
by Feigin \cite{Fei}. This cohomology is defined as
follows (see \cite{FFa,BMPb} for more details):
Introduce a ghost system $(b_A,c^A)$ for each generator $e_A$
of $\bfg$, with (anti-)commutators
$\{b_A,c^B\} = \de_A{}^B$, and denote the corresponding
ghost Fock space $F^{\gh}$. The (physical) ghost vacuum $|\gh\rangle$
satisfies
\eqn{eqBg}{
b_\al |\gh\rangle=b_i |\gh\rangle = c^{-\al} |\gh\rangle =0\,,\qquad
\al\in\De_+,\, i=1,\ldots,{\rm rank\,}\bfg\,.
}
The ghost Fock space is graded by ghost number, $\gh(c^A)=1,\,
\gh(b_A)=-1$, and is a $\bfg$-module under the action
\eqn{eqBi}{
\pi^{\gh}(e_A) = - \sum_{B,C} f_{AB}{}^C c^Bb_C\,.
}
Note that the highest weight of $F^{\gh}$ equals $2\rh$, where
$\rh$ is the principal vector of $\bfg$ ($(\rh,\al)=1,\,\forall\al\in\De_+$),
as is easily computed via
\eqn{eqBj}{
\pi^{\gh}(h_i)|\gh\rangle = -\sum_{\al\in\De_+} f_{i -\al}{}^{-\al}
c^{-\al}b_{-\al}|\gh\rangle = \sum_{\al\in\De_+} (\al_i^\vee,\al) |\gh\rangle
= (\al_i^\vee,2\rh) |\gh\rangle\,.
}
The (twisted) cohomology $\rmHtw(\bfg,V)$ is defined as
the cohomology of the (BRST) operator
\eqn{eqBh}{
d = \sum_A c^A \left( \pi(e_A) + \half \pi^{\gh}(e_A) \right) \,,
}
acting on the (graded) complex $V\otimes F^{\gh}$.

The twisted cohomology of a subalgebra of $\bfg$ is defined
similarly by restricting to the appropriate subset
of generators. In particular we are interested in the cohomologies
of (twisted) nilpotent subalgebras $\bfnp^w \equiv
w\cdot \bfnp\cdot w^{-1}$, corresponding to Weyl group elements $w\in W$.
In this case the sums run over $\al\in w(\De_+)$.

To orient the discussion it is worth noting in the ``untwisted'' case
($w'=1$) that
the computation of $\rmHtw^i(\bfnp,V)_\la \cong \rmH^i(\bfnp,V)_\la \cong
{\rm Ext}_{\cO}^i(M_\la,V)$ for various modules $V\in\cO$ is a classical
problem in mathematics. In the case when $V$ is a finite-dimensional
irreducible module $L_\La$ the result is well-known \cite{Kos}
(see (\ref{eqBc}) below, where this result is derived as an illustration
of standard techniques). For many other interesting modules, such as
Verma modules, \index{Verma module}
the problem has only been solved partially.

Besides Verma modules $M_\la$ and contragredient Verma modules
${\overline M}_\la$ there exists a class of modules in $\cO$,
the so-called twisted Fock spaces $F^w_\la$ (labelled by elements
$w\in W$), that interpolate between $M_\la$ and ${\overline M}_\la$.
[Our conventions are such that $F^1_\la \cong {\overline M}_\la$
and $F^{w_0}_\la \cong M_\la$.] These modules are,
in a sense, finite-dimensional analogues of Wakimoto modules \cite{Wak}
and were introduced in \cite{FFa}
(see \cite{BMPa} for explicit realizations).
They are uniquely characterized by the property that they are free
over $\cU(\bfnp^w \cap \bfnm)$, cofree over $\cU(\bfnp^w\cap\bfnp)$ and have
a unique highest weight vector (of weight $\la$).

The cohomology of a twisted Fock space $F^w_{\la}$ with respect to
the nilpotent subalgebra $\bfnp^w$ with the same twist $w$, is given by
\cite{FFa,BMPb}
\eqn{eqBa}{
\rmHtw^i(\bfnp^w,F^w_{\la}) \cong \de^{i,0}\ \CC_{\la+\rh-w\rh}\,.
}
In fact, (\ref{eqBa}) uniquely characterizes the module $F^w_{\la}$
in the category $\cO$.

For $\La$ an integral dominant weight, \ie $\La\in P_+$, there
exist resolutions \index{resolution}
of the irreducible module $L_\La$ in terms of twisted Fock spaces $F^w_\la$
(for any $w\in W$) with terms
\eqn{eqBb}{
C^i_w L_\La
  \cong \bigoplus_{\{ \si\in W | \ell_w(\si)=i\} } \ F^w_{\si*\La}\,,
}
where $\ell_w(\si)$ is the twisted length of $\si\in W$, which can be
expressed in terms of the usual length $\ell$ through
$\ell_w(\si) = \ell(w^{-1}\si) - \ell(w^{-1})$ and $\si*\La = \si(\La+\rh)
-\rh$ denotes a shifted action of the Weyl group.

To illustrate an application of (\ref{eqBb}) we reproduce the known
result for $\rmHtw(\bfnp^w,L_\La)$ as alluded to above.
Simply take a resolution of
$L_\La$ in terms of Fock spaces twisted by the same $w\in W$ and apply
(\ref{eqBa}) to the resulting double complex. We find
\eqn{eqBc}{
\rmHtw^i(\bfnp^w,L_\La) \cong \bigoplus_{\{ \si\in W | \ell_w(\si)=i\} }
\CC_{\,\si*\La + \rh -w\rh}\,.
}

We are interested in computing the twisted cohomology
$\rmHtw^i(\bfg,F^{w}_{\la} \otimes F^{w'w_0}_\mu)$, or rather the
cohomology relative to the Cartan subalgebra $\bft$, which we will denote
by $\rmHtw^i(\bfg,\bft;F^{w}_{\la} \otimes F^{w'w_0}_\mu)$.
[The Weyl group element
$w_0$ denotes the unique element of longest length in $W$.]
This cohomology corresponds to the physical states
\index{physical state} of the
finite-dimensional analogue of the so-called $G/G$-model.
Using the fact that $\bfg \cong \bfnp^{w'} \oplus \bft \oplus \bfnp^{w'w_0}$,
it can be related to a generalization of (\ref{eqBa}) by
invoking a reduction theorem (see \eg \cite{BMPb})
\eqna{eqBd}{
\rmHtw^i(\bfg,\bft;F^{w}_{\la} \otimes F^{w'w_0}_\mu) & \cong &
\bigoplus_{p+q=i} \left( \rmHtw^p(\bfnp^{w'},F^{w}_\la) \otimes
\rmHtw^q(\bfnp^{w'w_0},F^{w'w_0}_\mu) \right)_{\bft} \nonumber \\
& \cong & \rmHtw^i(\bfnp^{w'},F^{w}_\la)_{-\mu -\rh - w'\rh} \,.
}
One can show that nonzero cohomology can only arise if $\la$ and $\mu$
can be parametrized as
\eqn{eqBe}{
\la = \si*\La\,,\qquad\qquad \mu = -\si'*\La -2\rh\,,
}
for some dominant weight $\La$ and $\si\preceq\si'\,,\ \si,\si'\in W$.
[We take the usual Bruhat ordering ``$\preceq$'' on $W$ \cite{Hump}.
In particular for $sl(2)$ and $sl(3)$, this is just the ordering of
$W$ by the length $\ell(\cdot)$.]
Moreover, if one restricts the discussion to dominant integral weights,
\ie $\La\in P_+$, then one can show that the cohomology
does not depend on the particular $\La\in P_+$. We will
henceforth restrict the discussion to $\La\in P_+$ and adopt
the shorthand notations $F^{w}_\si = F^{w}_{\si*\La}$ and
$F^{w'w_0}_{-\si'} = F^{w'w_0}_{-\si'*\La - 2\rh}$.

In order to summarize the computation of the dimensions of
these cohomology groups, we introduce a set of polynomials by
\eqn{eqBf}{
\cP^{w,w'}_{\si,\si'}(q) = \sum_i (-1)^{\ell_w(\si)+\ell_w(\si')+i}\ q^i \
  {\rm dim\,} \rmHtw^i(\bfg,\bfh;F^{w}_\si \otimes F^{w'w_0}_{-\si'})\,.
}
They satisfy the following basic relations:
\begin{enumerate}
\item $\cP^{w,w'}_{\si,\si'}(q) = 0$ for $\si\succ\si'\,.$
\item  (i) $\ \cP^{w,w'}_{\si,\si}(q) =1\,,\quad$
(ii) $\ \cP^{w,w}_{\si,\si'}(q) = \de_{\si,\si'}\,,\quad$
(iii) $\ \cP^{w,w'}_{\si,\si'}(1) = \de_{\si,\si'}\,.$
\item $\cP^{w,w'}_{\si,\si'}(q) = \cP^{w'w_0,ww_0}_{\si' w_0,\si w_0}(q)
  \qquad$ (reflection symmetry).
\item $\cP^{w,w'}_{\si,\si'}(q) = \cP^{ww_0,w'w_0}_{\si,\si'}(q^{-1})
  \qquad$ (Poincar\'e duality).
\end{enumerate}

The identities in 2) follow from the fact that (i) for $\si=\si'$
there is only one state in the complex, (ii) for $w=w'$ the cohomology is
given by (\ref{eqBa}), and (iii) by applying the Lefschetz principle.
Identity 3) follows from the observation that in (\ref{eqBe}) we
could equally well have swapped the order of the two Fock spaces and
chosen the weight $\La'=-w_0\La$ to parametrize $(\la,\mu)$.
Finally, identity 4) follows from the fact that the module contragredient
to $F^w_\la$ is $F^{ww_0}_\la$.
In Appendix A we list all polynomials for $\bfg \cong sl(2)$ and $sl(3)$.

In the particular case $(w,w')=(w_0,1)$, where the polynomials correspond
to the $\bfg$ cohomology of a product of two Verma modules $M_\si
\otimes M_{-\si'}$, it is known that
$\cP^{w_0,1}_{\si,\si'}(q) = R_{\si,\si'}(q)$
for $\ell(\si') - \ell(\si) \leq3$ \cite{GaJo}. Here, $R_{\si,\si'}(q)$
denote the Kazhdan-Lusztig $R$-polynomials \cite{KaLu,Hump}.

The above discussion has a straightforward generalization to the
affine Lie algebras, in which case one is interested in computing
the relative semi-infinite cohomology
${\rmH}^{\infty/2+i}(\hg,\widehat \bft;F_{\la}^w\otimes
F_\mu^{w'w_0})$ of the tensor product of two Wakimoto
modules, twisted by finite-dimensional Weyl group
elements $w,w'\in W$ \cite{BMPb}. Once more the  cohomology can arise
only for the weights satisfying the affine analogue of
(\ref{eqBe}), with $\si$, $\si'$ in  the affine Weyl group $\wW$.

It is known that for $w=w'$ (see, \eg sections 4
and 5 in \cite{BMPb})
\eqn{eqKa}{
{\rmH}^{\infty/2+i}(\widehat \bfg,\widehat
\bft;F_{\si}^w\otimes F_{-\si'}^{ww_0})\cong \de_{\si,\si'}
  \de^{i,0} \ \CC\,,
}
which is the analogue of (\ref{eqBa}). We expect that for general
$w$ and $w'$ the set of polynomials as in (\ref{eqBf}) with
$\si\,,\si'\in\wW$ will in fact be the same as in the finite-dimensional
case. For $\widehat{sl}(2)$ this can be verified explicitly
from the results of \cite{Son}.
The general case appears to be an open problem.


\newsec{$\cW$-cohomology of Fock spaces at irrational $\al_+^2$}

Let $\cW[\bfg]$ be the $\cW$-algebra associated to some simply-laced simple
Lie algebra $\bfg$ (see \cite{BS} for a review and a list of notations).
In the remainder of this paper we will present some new results for the
semi-infinite
cohomology $\rmH^i(\cW[\bfg], V^M \otimes V^L)$ of $\cW[\bfg]$ on
the product of two positive energy $\cW[\bfg]$ modules $V^M$ and $V^L$.
Specifically, for the `Liouville' module $V^L$, representing the
$\cW[\bfg]$ gravity sector, we will take the Fock space of an appropriate set
of free scalar fields, while for the matter module $V^M$ we will take either
a Fock space (section 3) or an irreducible module, \ie $\cW[\bfg]$ minimal
model (section 4). Although, in general, the Cartan subalgebra of
$\cW[\bfg]$ will not be diagonalizable on the modules $V$, there is still
an analogue of the relative cohomology. One can show that the cohomology
possesses a multiplet structure of $2^\ell$ states (where $\ell= {\rm rank\,}
\bfg$), which is essentially due to the ghost
zero modes \cite{Pope,BMPc}.
The lowest ghost number state in each multiplet will be called
a prime state. Throughout this paper we will only formulate the results
for the prime states in the cohomology.

To be precise, our results are only valid for
$\bfg \cong sl(2)$ and $sl(3)$, for these are the only cases for which
the differential (BRST operator) has been constructed explicitly
(see \cite{TM,BLNWa} for the latter,
and also \cite{Zhu} for some higher rank results).
However, one expects
that such a differential exists for the other $\cW[\bfg]$ algebras. In
the discussion below we use only very generic properties of the
differential, and our results should therefore be valid for the other
$\cW[\bfg]$ algebras as well.

Let $F(\La,\al_0)$ denote the Fock space of $\ell \equiv {\rm rank\,}\bfg$
scalar fields $\ph^k(z)$, normalized such that $\ph^k(z) \ph^l(w)
= -\de^{kl} \ln(z-w)$, coupled to a background charge $\al_0\rh$.
The Fock space vacuum $|\La\rangle$ is labelled by a vector $\La$ in the
weight space of $\bfg$ such that $p^k|\La\rangle = \La^k |\La\rangle$.

A realization of $\cW[\bfg]$ on the Fock space $F(\La,\al_0)$
can be constructed
by means of the Drinfel'd-Sokolov reduction. In particular, the
stress energy tensor is given by
\eqn{eqCc}{
T(z) = - \half (\partial\ph(z)\cdot\partial \ph(z)) - i\al_0 \rh
  \cdot \partial^2\ph(z)\,.
}
It generates the Virasoro subalgebra of $\cW[\bfg]$. The central charge
and conformal dimension of $F(\La,\al_0)$ are given by
\eqn{eqCd}{
c = \ell - 12\al_0^2 |\rh|^2\,,\qquad h(\La) = \half (\La,\La+2\al_0\rh)\,.
}

Let us first consider the cohomology
$\rmH^i(\cW[\bfg],F(\La^M,\al_0^M) \otimes F(\La^L,\al_0^L))$.
Imposing the condition that the total central charge vanishes
in order for the differential to be nilpotent, leads to the following
parametrization of $\al_0^M$ and $\al_0^L$
\eqn{eqCy}{
\al_0^M = \al_+ + \al_-\,,\qquad -i\al_0^L = \al_+ - \al_-\,,\qquad
\al_+\al_- =-1\,.
}
By standard arguments, based on the composition series of a Fock space,
one can show that the cohomology is trivial (\ie
contains at most tachyonic states)
if either $F(\La^M,\al_0^M)$ or $F(\La^L,\al_0^L)$ is irreducible
(see \eg \cite{LZ}). Recall that a Fock
space $F(\La,\al_0)$ is reducible if and only if there exists
a root $\al\in\De_+$ such that (see \eg \cite{BMPc})
\eqn{eqCz}{
(\La+\al_0\rh,\al)\in \pm (\NN \al_+ + \NN \al_-)\,,
}
and `completely degenerate' if (\ref{eqCz}) holds for all roots $\al\in
\De_+$. Therefore, in the reducible case, it is convenient
to parametrize $\La$ by
\eqn{eqCe}{
\La + \al_0\rh  = w^{-1} (\al_+ \si (\Lap+\rh) + \al_-(\Lam+\rh))
  \equiv \La(w,\si) + \al_0\rh\,,
}
for some weights $\Lap,\Lam$ and $w,\si\in W$.
If $F(\La,\al_0)$ is reducible in the direction of $\al\in \De_+$,
then we can
choose $\Lap$ and $\Lam$ such that $(\Lap,\al)\in\NN,\,(\Lam,\al)\in\NN$.
If $F(\La,\al_0)$ is
completely degenerate then one can choose $\Lap,\Lam \in P_+$ as well
as $\si=1$, and if
moreover $\al_+^2$ is irrational then this parametrization of $\La$ in
terms of $\Lap,\Lam\in P_+$ and $w\in W$ is unique. In this paper we will
only consider the completely degenerate case. In the case that
$F(\La,\al_0)$ is reducible
only in certain directions the cohomology essentially reduces to the one
of a smaller $\cW$-algebra. It can be analysed similarly.

The highest weight vectors of all $F(\La(w,1)),\, w\in W$, have the
same eigenvalues with respect to the $\cW[\bfg]$ generators
because these are
invariant under $\La+\al_0\rh \to w( \La+\al_0\rh)$. Therefore,
all $F(\La(w,1))$ contain one and the same irreducible $\cW[\bfg]$
module $L(\La)$.
Thus, by arguing in the usual way,
we obtain a family of resolutions $C^i_wL(\La)$ of $L(\La)$,
parametrized by $w\in W$, that are based on $F(\La(w,1))$.
For $\al_+^2 \not\in \QQ$ these resolutions
have a finite number of terms (see section 4 for $\al_+^2\in\QQ$).
Specifically, for $\La = \al_+ \Lap + \al_-\Lam$  we have
\eqn{eqCf}{
C_w^iL(\Lap,\Lam) \ \cong\ \bigoplus_{
\{ \si\in W | \ell_w(\si)=i\} }\ F(\La(w,\si))\,.
}
where $\La(w,\si)$ is defined in (\ref{eqCe}). For $w=1$ these
resolutions were already constructed in \cite{Fre,Nie}.

\noindent Let us now turn to the discussion of the
cohomology
$\rmH^i(\cW[\bfg],F(\La^M,\al_0^M) \otimes F(\La^L,\al_0^L))$.
By analysis of the Kac-determinants for $\cW[\bfg]$ \cite{BMPc} and
standard reasoning using composition series for $F$,
one can argue that:\bigskip

\noindent{\it
For $\La^M$ of the form (\ref{eqCe}), \ie
\eqn{eqCh}{
\La^M(w,\si) +\al_0^M\rh = w^{-1}\left( \al_+ \si(\Lap + \rh) + \al_-
  (\Lam + \rh) \right) \,,
}
the cohomology
$\rmH^i(\cW[\bfg],F(\La^M,\al_0^M) \otimes F(\La^L,\al_0^L))$
can only be nontrivial if \hfill\\
$\La^L = \La^L(w',\si')$ for some $w',\si'\in W$,
where
\eqn{eqCi}{
-i(\La^L(w',\si') + \al_0^L \rh )  =  w'{}^{-1}\left(\al_+ \si' (\Lap + \rh)
  + \al_- (\Lam + \rh)\right)\,.
}}

At this point we note a remarkable similarity between the resolutions
of (\ref{eqCf}) and (\ref{eqBb}), which suggests that the matter module
$F(\La^M(w,\si))$ behaves like the twisted Wakimoto module $F^w_\si$,
while the Liouville module $F(\La^L(w',\si'))$ behaves like
the dual of the matter module, \ie like $F^{w'w_0}_{-\si'}$.
For irrational $\al_+^2$ we therefore expect a close correspondence
between the $\cW[\bfg]$ cohomology of a tensor product of two Fock
spaces and the (twisted) $\bfg$ cohomology of a tensor product of two
twisted Wakimoto modules. More precisely, we assert that
\eqn{eqCa}{
\sum_i (-1)^{\ell_w(\si) + \ell_{w'}(\si') + i}\ q^i\  {\rm dim\,}
\rmH^i(\cW[\bfg],F(\La^M(w,\si)) \otimes F(\La^L(w',\si'))) =
\cP^{w,w'}_{\si,\si'}(q)\,,
}
where the polynomials $\cP^{w,w'}_{\si,\si'}(q)$ were defined in
(\ref{eqBf}) (see Appendix A for an explicit list of all polynomials
for $\bfg \cong sl(2)$ and $sl(3)$).

For $\bfg \cong sl(2)$ the result agrees with \cite{LZ,BMPe}. For
$sl(3)$ the result (\ref{eqCa}) is in complete agreement with the states
explicitly constructed in \cite{BLNWb} (in \cite{BLNWb} only
states corresponding to, in our conventions, $w=\si,\, w'=\si'$ of
ghost number $\ell_{w'}(w') = - \ell(w')$ were considered).
We have explicitly constructed some additional physical states. The
results are consistent with (\ref{eqCa}).

Let us now examine the consequences of (\ref{eqCa}) for the
(chiral) ground ring \index{ground ring}
at irrational $\al_+^2$ in the $\cW[sl(3)]$ case.
{}From the explicit results in Appendix A we
conclude that for every $\Lap,\Lam\in P_+$ we have a ground
ring element iff $(w,w') = (1,w_0),\, (\si,\si') = (1,w_0)$. Moreover,
this element is unique. Let us denote it by $\ph_{(\Lap,\Lam)}$.
The ground ring is
freely generated by $\ph_{(\La_1,0)}, \ph_{(\La_2,0)}, \ph_{(0,\La_1)}$
and $\ph_{(0,\La_2)}$ (in the notation of \cite{BLNWb} these correspond
to $\ga_2^0, \ga_1^0, x_1$ and $x_2$, respectively).
[Note that our present conventions differ from those in \cite{BMPc}.]

It is reasonable to believe that an equation similar to
(\ref{eqCa}) holds for rational
$\al_+^2$  if one replaces the
$\bfg$ Fock space $F^w$ with its corresponding affinization \ie
a (twisted) Wakimoto module of $\hg$,
and let $\si$ run over the affine Weyl group. In particular we
claim (see (\ref{eqKa})):\bigskip

\noindent{\it
$\rmH^i(\cW[\bfg], F(\La^M(w,\si)) \otimes F(\La^L(w,\si')))$
(\ie $w=w'$) is nontrivial only if $\si=\si'$ in which case it is
one-dimensional and concentrated in dimension $i=0$ (tachyonic state).
}\bigskip

For $\widehat{sl}(2)$ this claim is known to be correct, while for other
$\hg$ it is consistent with sample calculations. In particular, this
assertion leads to the cohomology of $\cW$ minimal models (section 4)
where it does not contradict previous results.
[In the affine Lie algebra case it corresponds to (\ref{eqKa}).
For the analogous
statement for {(contragredient-)} Verma modules see \cite{BMPc}.]

For $\bfg \cong sl(2)$ it is well-known that the dimensions of the
cohomology groups
$\rmH^i(\cW[\bfg], F(\La^M(w,\si)) \otimes F(\La^L(w',\si')))$
are insensitive as to whether $\al_+^2$ is rational or irrational, as
a consequence of an $SO(2,\CC)$ symmetry relating different values of
$\al_+$ \cite{LZ}. This $SO(2,\CC)$ symmetry does not persist to
higher rank $\cW$-algebras, and in fact for $\bfg= sl(3)$ it has been
shown
that the cohomology at $c^M=2$ (corresponding to $\al_\pm = \pm1$) is
considerably larger than for irrational $\al_+^2$
(see \cite{BMPd} for some results for $\cW_3$ at $c^M=2$). There are two,
possibly related, reasons for this phenomenon. Firstly, for $\al_+^2
=\pm1$ the parametrization (\ref{eqCe}) in terms of $\Lap,\Lam\in P_+$ is
still highly redundant. Secondly, at $\al_\pm = \pm1$, both the complex and
the cohomology carry a representation of $\bfg$, where the $\bfg$-generators
are given by the (zero modes of a) Frenkel-Kac-Segal vertex operator
construction in the matter sector.


\newsec{$\cW$-cohomology of minimal models}

In this section we will give a complete classification of physical states for
a $\cW[\bfg]$ minimal model coupled to $\cW[\bfg]$ gravity.

The $\cW[\bfg]$ minimal models
arise for $\al_+^2 = {p/p'} \in \QQ$  ($p$ and $p'$ relatively prime integers)
and are labelled by two integrable weights
$\Lap\in P_+^{p-h^\vee}$ and $\Lam\in P_+^{p'-h^\vee}$ such that
the highest weight is given by $\La = \al_+\Lap + \al_- \Lam$.
Here, $P_+^k$ denotes the set of integrable weights of $\hg$ at level $k$
and $h^\vee$ is the dual Coxeter number.

We have a set of resolutions, parametrized by $w\in W$,
of the minimal model $L(\Lap,\Lam)$ in
terms of twisted Fock spaces similar to (\ref{eqCf})
\eqn{eqDc}{
C^i_wL(\Lap,\Lam)\ \cong\ \bigoplus_{
  \{ \si\in \wW | \ell_w(\si)=i\} }\ F(\La(w,\si))\,,
}
where $\La(w,\si)$ is defined as in (\ref{eqCe}), but now the sum over $\si$
runs over the affine Weyl group $\wW$. [We recall that any $w\in \wW$ can
be written as $w= t_\be \bar{w}$ for some $\bar{w}\in W$ and translation
$t_\be$ such that $w\la = t_\be \bar{w} \la = \bar{w}\la + k \be$ for
affine weights $\la$ of level $k$. In the following $\rh$ is regarded
as an element of $P_+^{h^\vee}$.]
The twisted length $\ell_w$ on the affine Weyl group $\wW$ is
defined by (see \cite{FFa,BMPa})
\eqn{eqDd}{
\ell_w(\si) = \lim_{N\rightarrow \infty} \left(\ell(t_{-Nw\rh} \si )-
  \ell(t_{-Nw\rh})\right)\,.
}
Since, for each $\si\in \wW$ there are only a finite number of possible
cancellations between $t_{-Nw\rh}= wt_{-N\rh}w^{-1}$ and $\si$, the limit in
(\ref{eqDd}) is in fact reached at a finite value of $N$.
Note furthermore that for $\si\in W$ the length defined
by (\ref{eqDd}) reduces to the usual
twisted length $\ell_w(\si)=\ell(w^{-1} \si) - \ell(w^{-1})$.

Now, consider the cohomology $\rmH^i(\cW[\bfg],
L(\Lap,\Lam)\otimes F(\La^L,\al_0^L))$.
By taking an arbitrary resolution $C^i_wL(\Lap,\Lam)$
of $L(\Lap,\Lam)$ one finds that the cohomology is
nontrivial if and only if $\La^L = \La^L(w,\si)$ for some $w\in W$
and $\si\in \wW$ where
\eqn{eqDf}{
 -i(\La^L(w,\si) + \al_0^L \rh) = w^{-1}\left( \al_+\si(\Lap+\rh) +
  \al_-(\Lam+\rh) \right)  \,.
}
Now, as discussed in section 3,
$\rmH^i(\cW[\bfg], F(\La^M(w,\si)) \otimes F(\La^L(w,\si'))) \cong
\de_{\si,\si'}\de^{i,0}\CC$. Thus by taking, for any $\La^L
= \La^L(w,\si)$, a resolution $C^i_wL(\Lap,\Lam)$ of
$L(\Lap,\Lam)$ with the same twist $w\in W$, the same argument as in \eg
\cite{BMPe} immediately yields\bigskip

\noindent{\it
1.\  The cohomology $\rmH^i(\cW[\bfg], L(\Lap,\Lam)
\otimes F(\La^L,\al_0^L))$ is nontrivial iff $\La^L=\La^L(w,\si)$ for
some $w\in W$ and $\si\in \wW$.\hfill\\
2.\ For $\La^L=\La^L(w,\si)$ there is precisely one (prime) state
in the cohomology. Its ghost number is given by $\ell_w(\si)$ and
its energy level by $E = \half |\al_+(\Lap +\rh)+ \al_-(\Lam+\rh)|^2
- \half |\La^L(w,\si)+\al_0^L\rh|^2$.
}\bigskip

In \cite{BMPc} we obtained this result for
the particular case that $-i(\La^L +
\al_0^L \rh)$ is in the fundamental Weyl chamber $D_+$. This corresponds
to those $\La^L(w,\si)$
where for any $\si\in\wW$ the Weyl group element $w=w_\si$ is determined
in such a way that $ -i(\La^L(w_\si,\si) + \al_0^L \rh) \in D_+$.
The above result extends our previous work
to all Weyl chambers. For $\bfg\cong sl(2)$ it agrees with \cite{LZ,BMPe}.

For $\bfg\cong sl(3)$ and the trivial module $L(\Lap,\Lam)\cong \CC$
(\ie $p=3,\,p'=4$ and $\Lap=\Lam=0$) the results can be compared to those
for the `two-scalar $\cW_3$ string' \cite{Pope}. We find a complete
agreement. For illustrative purposes we provide, in this case, a
table of physical states for low ghost numbers.
In each row the table lists, for given $\si\in\wW$ ($\ell(\si) \leq4$),
the values of the (energy-)level $E$ and
the ghost number $\ell_w(\si)\,,\ w\in W$
(see (\ref{eqDd})), of the corresponding (prime) state
in $\rmH^i(\cW_3, \CC \otimes F(\La^L(w,\si)))$. The table can be
compared to the results of \cite{Pope}.

\bigskip

\noindent{\bf Acknowledgements:} P.B.\ would like to
thank the organizers of the Carg\`ese workshop for the invitation and
opportunity to present this talk.
J.M.\ would like to thank U.S.C., and
P.B.\ and K.P.\ the University of Adelaide, for hospitality.
We enjoyed several discussions with Robert Perret on the
$\bfg$ cohomology.
This work was supported by the
Packard Foundation, by the U.S.\
Department of Energy under contract \#DE-FG03-84ER-40168 and
the Australian Research Council.

\setcounter{section}{0}
\newapp{\bf Appendix}

In this appendix we give some
explicit polynomials $\cP^{w,w'}_{\si,\si'}(q)$. In particular
we present an exhaustive list for $\bfg\cong sl(2)$ and $sl(3)$.
The results have been verified by explicit computation.

\eqn{eqEa}{
\cP^{w,w}_{\si,\si'}(q) = \cases{
  1 & if $\si=\si'$ \cr
  0 & otherwise \cr}
}
\eqn{eqEb}{
\cP^{w,wr_i}_{\si,\si'}(q) = \cases{
  1 & if $\si=\si'$ \cr
  q^{\pm1}-1 & if $ \si \prec r_{w\al_i} \si =\si',\, w\al_i\in\De_\mp$ \cr
  0 & otherwise \cr}
}
For $i\neq j$
\eqn{eqEc}{
\cP^{w,wr_ir_j}_{\si,\si'}(q) = \cases{
  1 & if $\si=\si'$ \cr
  q^{\pm1} - 1 & if $\si \prec r_{w\al_i}\si = \si'\,,\ w\al_i\in \De_\mp$ \cr
          & or $\si \prec r_{wr_i\al_j}\si=\si'\,,\ wr_i\al_j\in\De_\mp$ \cr
  (q^{\pm1} -1)^2 & if $\si \prec r_{w\al_i} \si
     \prec r_{w\al_i}r_{w\al_j} \si =\si'   \,,\ w\al_i\in\De_\mp\,,$\cr
      & $wr_i\al_j \in \De_\mp$ \cr
  (q-1)(q^{-1} -1) & if $\si \prec r_{w\al_i} \si
     \prec r_{w\al_i}r_{w\al_j} \si =\si'   \,,\ w\al_i\in\De_\mp\,,$\cr
     & $wr_i\al_j \in \De_\pm$ \cr}
}

Note that (\ref{eqEc}) can be summarized as
\eqn{eqEd}{
\cP^{w,wr_ir_j}_{\si,\si'}(q) =
  \sum_{\si\preceq\si''\preceq\si'} \cP^{w,wr_i}_{\si,\si''}(q)
  \cP^{wr_i,wr_ir_j}_{\si'',\si'}(q)\,.
}
The right hand side of (\ref{eqEd}) is in fact the polynomial associated
to the first term in the spectral sequence of $\rmHtw^i(\bfg,\bft;
F^{w}_{\si} \otimes F^{wr_ir_jw_0}_{-\si'})$ with respect to the
decomposition $\bfg \cong \bfnp^{wr_i} \oplus \bft \oplus \bfnp^{wr_iw_0}$.
In this particular case this spectral sequence collapses at the first term,
hence the result (\ref{eqEd}).

For $sl(3)$ the above determines all but the six polynomials
$\cP^{w,ww_0}_{\si,\si'}(q)$, which are listed below

\eqn{eqEe}{
\cP^{w_0,1}_{\si,\si'}(q) = \cP^{1,w_0}_{\si,\si'}(q^{-1}) =\cases{
  1 & if $\si=\si'$ \cr
  q-1 & if $\ell(\si') - \ell(\si) = 1$ \cr
  (q-1)^2 & if $\ell(\si') - \ell(\si) = 2$ \cr
  (q^3 - 2q^2 + 2q -1) & if $\ell(\si')-\ell(\si) =3$ \cr}
}
For $i,j\in \{1,2\},\, i\neq j$
\eqn{eqEf}{
\cP^{r_i,r_jr_i}_{\si,\si'}(q) = \cP^{r_jr_i,r_i}_{\si,\si'}(q^{-1}) = \cases{
  1 & if $\si=\si'$ \cr
  q - 1 & if $\si\prec r_i\si = \si' $ \cr
        & or $\si\prec r_{r_j\al_i}\si=\si'$ \cr
  q^{-1}-1 & if $\si\prec r_j\si=\si'$ \cr
  (q-1)(q^{-1}-1) & if $\si\prec \si r_i \prec \si r_ir_j =\si'$ \cr}
}



\begin{table} \centering
\begin{tabular}{|| l || r || r | r |    r | r |  r | r ||}\hline
 $\sigma\backslash w$ & E & $1$ & $r_1$ & $r_2$ & $r_2r_1$ &
 $r_1r_2$ & $r_1r_2r_1$ \\ \hline\hline

  $1$ & $ 0$ & $0$ & $0$ & $0$ & $0$ & $0$ & $0$  \\ \hline

  $r_1$ & $ 1$ & $1$ & $-1$ & $1$ & $1$ & $-1$ & $-1$ \\
  $r_2$ & $1 $ & $1$ & $1$ & $-1$ & $-1$ & $1$ & $-1$ \\
  $r_0$ & $ 2$ & $-1$ & $-1$ & $-1$ & $1$ & $1$ & $1$ \\ \hline

  $r_1 r_2$ & $ 3$ & $2$ & $0$ & $2$ & $0$ & $-2$ & $-2$ \\
  $r_2 r_1$ & $ 3$ & $2$ & $2$ & $0$ & $-2$ & $0$ & $-2$  \\
  $r_1 r_0$ & $ 4$ & $0$ & $-2$ & $2$ & $2$ & $-2$ & $0$ \\
  $r_2 r_0$ & $ 4$ & $0$ & $2$ & $-2$ & $-2$ & $2$ & $0$ \\
  $r_0 r_1$ & $ 5$ & $-2$ & $-2$ & $0$ & $2$ & $0$ & $2$ \\
  $r_0 r_2$ & $ 5$ & $-2$ & $0$ & $-2$ & $0$ & $2$ & $2$ \\
\hline

  $r_1 r_2 r_1$ & $ 4$ & $3$ & $1$ & $1$ & $-1$ & $-1$ & $-3$ \\
  $r_0 r_1 r_0$ & $ 6$ & $-1$ & $-3$ & $1$ & $3$ & $-1$ & $1$ \\
  $r_0 r_2 r_0$ & $ 6$ & $-1$ & $1$ & $-3$ & $-1$ & $3$ & $1$ \\
  $r_1 r_2 r_0$ & $ 8$ & $3$ & $-1$ & $3$ & $1$ & $-3$ & $-3$ \\
  $r_2 r_1 r_0$ & $ 8$ & $3$ & $3$ & $-1$ & $-3$ & $1$ & $-3$  \\
  $r_1 r_0 r_2$ & $ 9$ & $1$ & $-3$ & $3$ & $3$ & $-3$ & $-1$ \\
  $r_2 r_0 r_1$ & $ 9$ & $1$ & $3$ & $-3$ & $-3$ & $3$ & $-1$ \\
  $r_0 r_1 r_2$ & $ 11$ & $-3$ & $-3$ & $-1$ & $3$ & $1$ & $3$ \\
  $r_0 r_2 r_1$ & $ 11$ & $-3$ & $-1$ & $-3$ & $1$ & $3$ & $3$ \\
\hline

  $r_1 r_2 r_1 r_0$ & $ 10$ & $4$ & $2$ & $2$ & $-2$ & $-2$ & $-4$ \\
  $r_1 r_0 r_2 r_0$ & $ 11$ & $2$ & $-2$ & $4$ & $2$ & $-4$ & $-2$ \\
  $r_2 r_0 r_1 r_0$ & $ 11$ & $2$ & $4$ & $-2$ & $-4$ & $2$ & $-2$ \\
  $r_0 r_1 r_0 r_2$ & $ 13$ & $-2$ & $-4$ & $2$ & $4$ & $-2$ & $2$ \\
  $r_0 r_2 r_0 r_1$ & $ 13$ & $-2$ & $2$ & $-4$ & $-2$ & $4$ & $2$ \\
  $r_1 r_2 r_0 r_1$ & $ 14$ & $4$ & $0$ & $4$ & $0$ & $-4$ & $-4$ \\
  $r_2 r_1 r_0 r_2$ & $ 14$ & $4$ & $4$ & $0$ & $-4$ & $0$ & $-4$  \\
  $r_0 r_1 r_2 r_1$ & $ 14$ & $-4$ & $-2$ & $-2$ & $2$ & $2$ & $4$ \\
  $r_1 r_0 r_2 r_1$ & $ 16$ & $0$ & $-4$ & $4$ & $4$ & $-4$ & $0$ \\
  $r_2 r_0 r_1 r_2$ & $ 16$ & $0$ & $4$ & $-4$ & $-4$ & $4$ & $0$ \\
  $r_0 r_1 r_2 r_0$ & $ 18$ & $-4$ & $-4$ & $0$ & $4$ & $0$ & $4$ \\
  $r_0 r_2 r_1 r_0$ & $ 18$ & $-4$ & $0$ & $-4$ & $0$ & $4$ & $4$ \\
\hline
\end{tabular}
\caption{Physical states for the two-scalar $\cW_3$ string}
\end{table}


\end{document}